\documentclass{aastex631}

\usepackage{amssymb,amsmath}
\usepackage{graphicx,times}
\submitjournal{Res.\ Astron.\ Astrophys.} 
\accepted{October 8, 2021}

\shorttitle{Long-Term Evolution of AR NOAA 12673}
\shortauthors{Muhamad et al.}
\graphicspath{{./}}
\begin{document}

\title{Long-Term Evolution of Magnetic Fields in Flaring Active Region NOAA 12673}

\correspondingauthor{Johan Muhamad}
\email{johan.muhamad@lapan.go.id}

\author{Johan Muhamad}
\affiliation{Space Science Center, \\
National Institute of Aeronautics and Space (LAPAN), \\
National Research and Innovation Agency (BRIN), \\
Bandung 40173, Indonesia}

\author{Muhamad Zamzam Nurzaman}
\affiliation{Space Science Center, \\
National Institute of Aeronautics and Space (LAPAN), \\
National Research and Innovation Agency (BRIN), \\
Bandung 40173, Indonesia}

\author{Tiar Dani}
\affiliation{Space Science Center, \\
National Institute of Aeronautics and Space (LAPAN), \\
National Research and Innovation Agency (BRIN), \\
Bandung 40173, Indonesia}

\author{Arun Relung Pamutri}
\affiliation{Department of Mathematics, \\
Universitas Jenderal Soedirman, \\
Purwokerto, Indonesia}
             
%% Note that the \and command from previous versions of AASTeX is now
%% depreciated in this version as it is no longer necessary. AASTeX 
%% automatically takes care of all commas and "and"s between authors names.

%% AASTeX 6.31 has the new \collaboration and \nocollaboration commands to
%% provide the collaboration status of a group of authors. These commands 
%% can be used either before or after the list of corresponding authors. The
%% argument for \collaboration is the collaboration identifier. Authors are
%% encouraged to surround collaboration identifiers with ()s. The 
%% \nocollaboration command takes no argument and exists to indicate that
%% the nearby authors are not part of surrounding collaborations.

%% Mark off the abstract in the ``abstract'' environment. 
\begin{abstract}

During the lifetime of AR 12673, its magnetic field evolved drastically and produced numerous large flares. In this study, using full maps of the Sun observed by the Solar Dynamics Observatory and the Solar Terrestrial Relations Observatory, we identified that AR 12673 emerged in decayed AR 12665, which had survived for two solar rotations. Although both ARs emerged at the same location, they possessed different characteristics and different flare productivities. Therefore, it is important to study the long-term magnetic evolution of both ARs to identify the distinguishing characteristics of an AR that can produce large solar flares. We used the Spaceweather Helioseismic and Magnetic Imager Active Region Patch data to investigate the evolution of the photospheric magnetic field and other physical properties of the recurring ARs during five Carrington rotations. All these investigated parameters dynamically evolved through a series of solar rotations. We compared the long-term evolution of AR 12665 and AR 12673 to understand the differences in their flare-producing properties. We also studied the relation of the long-term evolution of these ARs with the presence of active longitude. We found that the magnetic flux and complexity of AR 12673 developed much faster than those of AR 12665. Our results confirmed that a strong emerging flux that emerged in the pre-existing AR near the active longitude created a very strong and complex AR that produced large flares.     

\end{abstract}

%% Keywords should appear after the \end{abstract} command. 
%% The AAS Journals now uses Unified Astronomy Thesaurus concepts:
%% https://astrothesaurus.org
%% You will be asked to selected these concepts during the submission process
%% but this old "keyword" functionality is maintained in case authors want
%% to include these concepts in their preprints.
%% From the front matter, we move on to the body of the paper.
%% Sections are demarcated by \section and \subsection, respectively.
%% Observe the use of the LaTeX \label
%% command after the \subsection to give a symbolic KEY to the
%% subsection for cross-referencing in a \ref command.
%% You can use LaTeX's \ref and \label commands to keep track of
%% cross-references to sections, equations, tables, and figures.
%% That way, if you change the order of any elements, LaTeX will
%% automatically renumber them.
%%
%% We recommend that authors also use the natbib \citep
%% and \citet commands to identify citations.  The citations are
%% tied to the reference list via symbolic KEYs. The KEY corresponds
%% to the KEY in the \bibitem in the reference list below. 

\section{Introduction}           %% first-level sections will be auto-capitalized
\label{sect:intro}

Majority of the existing solar flare prediction methods are dependent on short-term (several hours to days) physical properties of an active region (AR) and its flaring history \citep{Campi_2019, Leka_2019, Chen2019} although AR lifetimes can reach up to 10 and 5 months during the solar minimum and maximum, respectively \citep{Schrijver1994,vdg2015}. During an AR lifetime, its magnetic field properties determine the characteristics of its flare productivity and eruptivity. A previous study \citep{1999ASPC..184..302V} found that the long-lived AR 7978, which emerged in the late phase of solar cycle 22, survived for several months, became flare-rich in the first three rotations itself, and remained eruptive until the decay phase. \cite{Schrijver1994} found that the area of a bipole AR is proportional to the magnetic flux emerged in it during its lifetime. \cite{vdg2015} showed that an AR lifetime is proportional to the amount of magnetic flux at its maximum development. \cite{Ugarte_Urra_2015} found the lifetimes of several ARs to be continuously proportional to the maximum 304 $\mathrm{\AA}$ intensity. Recently, \cite{IGLESIAS20201641} studied long-term evolution of a long-duration AR in the early phase of solar cycle 24 and found that its characteristics varied significantly during five solar rotations. 

Several studies found the existence of a 25.5--27 day flare periodicity and its harmonics in the Sun, which were attributed to the solar rotation \citep{Bai_2003b,kilcik2010}. The fundamental $\approx$27 days of periodicity can be considered as an indicator of the existence of a long-lifetime sunspot or an AR that re-occurs in the Earth-side solar disk after one solar rotation. Furthermore, many studies argued that there exist several locations in the Sun where sunspots tend to emerge repeatedly during one solar cycle. The longitudes where sunspots or active regions are commonly observed over a long period are known as active longitudes \citep{Ivanov_2007,Gyenge_2016}. Because the active zones in the Sun do not pose a common longitude in the northern and southern hemisphere, the term “hotspot” is sometimes preferred to be used \citep{Bai_90,Bai_2003a}. The presence of active longitudes or hotspots might contribute to the periodicities of flare activities \citep{Gyenge_2016}. 
 
Based on the analysis of sunspot area over solar cycle 12--23, it was found that the active longitudes were distributed at near 90$^{\circ}$, 180$^{\circ}$, and 270$^{\circ}$ of Carrington longitudes in the northern hemisphere, and at 90$^{\circ}$, 180$^{\circ}$, and 360$^{\circ}$ of Carrington longitudes in the southern hemisphere \citep{Ivanov_2007}. It was also found that the width of the sunspot formation zone was approximately 20$^{\circ}$--30$^{\circ}$ heliongitude and rigidly rotated with a Carrington period. This sunspot formation zone could survive for 15--20 rotations. Moreover, previous studies revealed there were two active longitudes of sunspots that were roughly separated by 180 degrees over a long period of time \citep{Usoskin_2005}. The locations of active longitudes in both hemispheres were found to be asymmetry. This asymmetry was probably related to the different rotation rates that were observed in the northern and southern hemispheres \citep{Usoskin_2005,Zhang_2011}.  

Magnetic activity complex, where new magnetic flux emerges in the pre-existing AR, was also often observed in the Sun over multiple rotations \citep{Bumba_69, Gaizauskas_1983}. This newly emerging flux in pre-existing AR could increase the non-potentiality of new AR that led to the occurrences of solar flares \citep{Komm_2015}. \citet{Gyenge_2016} found ARs that were near active longitude had a higher probability to produce flares than ARs that were far from active longitude.
 
Owing to the relation of flare periodicities with solar rotation and active longitudes, the long-term (27 days) characteristics of a flaring AR may be useful for long-term flare prediction and understanding the process in the interior of the Sun. Thus, it is interesting to investigate whether the recent flare prediction capability can be improved. However, the long-term evolution of flaring ARs having long lifetimes in solar cycle 24 have not been extensively studied yet, although many similar studies have been conducted on ARs in past solar cycles. To explore this issue, we investigated the long evolution of an AR that produced many flares in solar cycle 24. We used AR 12673 as a representative of the flaring AR owing to its rich flaring history.  

The X9.3 solar flare which occurred on September 6, 2017 from AR NOAA 12673 was the most powerful flare during solar cycle 24. During the flaring period from September 4 to September 10, 2017 many other large flares and coronal mass ejections (CMEs) occurred from the same AR. Numerous severe impacts of these events on radio communications were reported \citep{redmon2018}. Because the energy and onset of a solar flare are related to the evolution of the magnetic field of the AR, it is particularly important to understand the flare productivity of AR 12673 by studying its magnetic field evolution over long period. 

Many studies have focused on understanding mechanism the energy build-up \citep{Yang_2017,Verma2018,Liu_2018} and flare trigger mechanism of AR 12673 that caused the numerous large flares and CMEs \citep{Yan_2018,Jiang_2018,ino18,Bamba_2020}. Analysis of photospheric magnetogram data and coronal nonlinear force-free field (NLFFF) extrapolations suggested that the large flares and CMEs in this AR were generated by combined shearing motion, sunspot rotation in the photosphere, and interaction of the magnetic flux rope and the surrounding field in the corona \citep{Liu_2018, Hou_2018} due to some intrusion of the opposite field in the vicinity of the polarity inversion line (PIL) \citep{Bamba_2020}. Photospheric magnetogram data have clarified that the highly dynamic in photosphere and sub-photosphere of AR 12673 caused a rapid helicity injection in the corona and was responsible for the eruptive flares that occurred on September 6, 2017 \citep{Vemareddy_2019,Moraitis2019}. All the above studies focused on the time evolution of the AR several days prior and after the X9.3 flare occurrence. We proposed that it is also necessary to study the long-time evolution of this AR because of the lack of clear evidence that previous ARs have been observed from the same region as that of AR 12673.

We traced the location of this AR for several rotations before and after it transited in the Earth-side solar disk. We found several NOAA ARs that preceded and succeeded AR 12673 at the same location. The long-term recurrence of these ARs at the same location allowed investigation of the long-term evolution of the magnetic field. For clarity, in our paper, we have used the terminology reoccurring AR to refer to a bundle of strong magnetic flux regions that emerged at similar locations over multiple solar rotations. A strong magnetic flux region appeared in the Sun in a specific solar rotation and identified as an AR without unique number assigned by the NOAA is called as AR NOAA (ARN). 

In this study, we aimed to show the photospheric magnetic field evolution of a flaring AR over multiple solar rotations. The objective of this study was to understand the relation between flaring AR 12673 and other ARs that were observed in the same region in several solar rotations before and after the passage of AR 12673. Our main goal was to identify the characteristics of the AR magnetic field during its lifetime over several rotations and its relation with flare activities. We wanted to know whether the flare periodicity can be used to improve the flare prediction time. In this study, we only considered two flaring ARNs as case studies. Therefore, the results may not represent the general characteristics of a long-lived AR evolution in the Sun for solar cycle 24. Nevertheless, this investigation can contribute to the studies on the long-term evolution of the AR 12673 magnetic field as well as long-term flare forecasting.

\section{Data and Methods}
\subsection{Data}

Photospheric magnetic field data were obtained using the Helioseismic and Magnetic Imager (HMI) on board the Solar Dynamics Observatory (SDO) that continuously measure the full-disk photospheric vector magnetic field every 720 seconds \citep{Schou_2012}. The inversion of the HMI Stokes I, Q, U, and V data are conducted by using the Very Fast Inversion of the Stokes Vector (VFISV) code, which assumes a Milne-Eddington model of the solar atmosphere \citep{Borrero_2011}. The 180° ambiguity of HMI vector magnetic field is resolved using a minimum energy method \citep{Metcalf_1994, Leka_2009}. The detailed pipeline processing of HMI vector magnetic field data is given by \citet{Hoeksema_2014}. 

Space-weather HMI Active Region Patches (SHARPs) is a derivative data product of HMI, which is released by the HMI team \citep{Bobra_2014}. It automatically identifies active region patches in HMI data and remaps the vector magnetic field data to a Lambert cylindrical equal-area projection (CEA) \citep{Bobra_2014}. We used collections of magnetogram data series of SHARPs, which contained three vector magnetic field components ($B_\phi, B_\theta, B_r$) (\url{http://jsoc.stanford.edu/doc/data/hmi/sharp/sharp.htm}). As shown by \citet{Bobra_2014}, the signal-to-noise ratio in the SHARPs parameters increases significantly beyond +/- 70$^{\circ}$ of central meridian. Because of the limitations in using the SHARP data for the location of an AR far from the disk center, we only used them during the passage of the ARs for four days of its closest position to the disk center. Therefore, it should be noted that this constraint hinders the application of our results to the condition of the AR at other longitudes. 

Synoptic maps were acquired from the HMI magnetic field synoptic charts for various Carrington rotations (CRs), which were also available at the Joint Science Operations Center (JSOC), Stanford University. We chose radial magnetic field data for the CRs 2192--2196, as shown in Figure 1. The flare histories and the magnetic classes were obtained from the Heliophysics Integrated Observatory (\url{https://helio-vo.eu}) catalogue. We defined the solar flare histories only for the C-class or stronger flares recorded in the abovementioned catalogue. We also used GOES X-ray flux data to study the periodicity of flare occurrence in 2017. The data were accessed from \url{https://satdat.ngdc.noaa.gov/sem/goes/data/}. We used a one-minute time resolution for the GOES-15 X-ray flux, complemented by the GOES-13 data to compensate the unavailable data from July to November 2017. When no data were recorded by the two satellites, we conducted a linear interpolation to fill the gaps. Full-disk HMI solar magnetogram data obtained from the Virtual Solar Observatory (\url{https://sdac.virtualsolar.org/}) were accessed and processed using Sunpy modules \citep{Mumford_2015}. 

To identify the Carrington longitudes of the AR for several solar rotations, we estimated the coordinates of the AR by considering the differential rotation rate at the observed latitudes. The synodic rotation rate was calculated using formula \citep{sheeley_1992}:
\begin{equation}\label{eq01}
\omega(\theta) = 13.46 - 2.7\cos^2 \theta + 1.2\cos^4 \theta - 3.2\cos^6 \theta   ,
\end{equation}
where $\theta$ is colatitude.

The Carrington longitudes of the ARNs in multiple rotations are listed in table 1. The estimated Carrington longitude was calculated for every CR using the rotation rate at the corresponding colatitude when the ARN was near the central meridian as a reference for the coordinate of the following CR. We compared the results from equation \eqref{eq01} with the Carrington coordinates of the ARNs given by the Heliophysics Event Catalogue (\url{https://helio-vo.eu/solar_activity/arstats-archive/}).

Data from the Atmospheric Imaging Assembly (AIA) instrument onboard the SDO and the Extreme Ultraviolet Imager (EUVI) onboard the STEREO were used to trace the ARs from July to November 2017. By combining the observations from the SDO/AIA at 193 $\mathrm{\AA}$ and the STEREO-A/EUVI at 195 $\mathrm{\AA}$ we produced a series of global maps of the Sun for five solar rotations. However, STEREO-B could not observe the Sun since 2014; therefore, complete views of the Sun in the Earth and far sides were impossible to obtain during this period. We traced AR 12665 from July 6 to November 3, 2017. First, we marked the location of the centroid of the bright region in AR 12665 on July 6 on the map and defined the location in the heliographic Stonyhurst longitude and latitude. Subsequently, we tracked this bright region by predicting its future longitude. The traced region in the fullmap of the Sun is shown as a red square in Figure 2. The prediction assumed that the bright region rotated at a certain rate. \cite{sharma2020} observed the rotation rate of the Sun in the AIA 193 $\mathrm{\AA}$ wavelength at a low latitude (-10$^{\circ}$ S to 10$^{\circ}$ N) as approximately 26--27$^{\circ}$/day. Note that this rotation rate is slower than that on the photosphere.                 

\subsection{Magnetic Free Energy}
We calculated the magnetic potentials of the ARs from the radial components of the magnetic fields of SHARP data using the Fourier method \citep{alissandrakis1981}. The magnetic potentials were calculated every step with a 3-hour time cadence. 

Magnetic energy of the ARs were calculated as follows:

\begin{equation}\label{eq1}
E = \int \frac{B^2}{8\pi} dV  ,
\end{equation}   
where E is the energy, B is the magnetic flux density, and V is the volume of the space over the AR.   

Because information was only available from two-dimensional photospheric magnetic field data, only the proxy of the volumetric magnetic energy can be directly calculated. The real free energy, $E_{rf}$ , of the AR in the photosphere can be defined as 
\begin{equation}\label{eq5}
E_{rf} = \frac{(\bf{B_{ob}}^2 - \bf{B_{pot}}^2)}{8\pi} dA  .
\end{equation}   

However, using equation \ref{eq5}, a negative free energy density may be obtained because $\bf{B_{pot}}^2$ may be greater than $\bf{B_{ob}}^2$ \citep{Zhang2016}. Therefore, instead of using equation \ref{eq5} to express the energy evolution of the AR in the photosphere, we preferred to calculate the nonpotential energy using the following formula:

\begin{equation}\label{eq2}
E_{np} = \int \frac{B_s ^2}{8\pi} dA   ,
\end{equation}     			
where 
\begin{equation}\label{eq3}
\bf{B_s} = \bf{B_{ob}} - \bf{B_{pot}}  .
\end{equation}   
$\mathbf{B_s}$ denotes the source field obtained from the difference between the observed field ($\mathbf{B_{ob}}$) and the potential field ($\mathbf{B_{pot}}$) \citep{yang2012, Zhang2016}. 

This source field represents the nonpotentiality of the magnetic field on the photosphere. It is noteworthy that these properties were only calculated when the ARNs were located near the disk center. In this study, to evaluate the free energy evolution of the ARs during multiple solar rotations, the free energy definition by \cite{yang2012} as expressed in equation \ref{eq2} was adopted. Thus, the free energy, $E_f$, was derived as
\begin{equation}\label{eq4}
E_f = \frac{(B_{ob} - B_{pot})^2}{8\pi} + \frac{B_{ob}B_{pot}}{2\pi} sin^{2}(\frac{\theta_{s}}{2}) dA.
\end{equation}
$\theta_{s}$ is the shear angle, which is the angle between the projected and potential vector magnetic fields on the photosphere. 
This definition is highly convenient for obtaining the temporal magnetic field evolution because it ensures a positive free energy density. 
   
To validate the calculation of the magnetic free energy on the photosphere, we compared the free energy calculated using equation \ref{eq4} with the volumetric free energy calculated using the NLFFF model. We employed the magnetohydrodynamics (MHD) relaxation method by \cite{ino14a} to extrapolate the NLFFF model of the AR. First, the three-dimensional (3-D) potential field of the AR was extrapolated using the Fourier method \citep{alissandrakis1981} based on the radial components ($B_r$) in the vector magnetic field data. Subsequently, the horizontal components ($B_{xp}$ and $B_{yp}$) on the bottom boundary of the potential field model were incrementally changed to the observed horizontal components ($B_{x}$ and $B_{y}$). The induction equation was solved iteratively until the bottom boundaries of the NLFFF satisfied the observed vector magnetic field components. For the detailed MHD relaxation method, see \cite{ino14a}. 

This NLFFF method has been evaluated to reproduce magnetic field of ideal force-free cases, i.e., Low and Lou solution, with high accuracy \citep{ino14a}. The method has been applied to produce NLFFFs of many famous ARs, e.g., AR 10930 \citep{ino11,ino12}, AR 11158 \citep{ino13,ino14b}, AR 12192 \citep{ino16b,bam17}, that were topologically comparable to the coronal fields of the ARs observed by SDO/AIA. For modeling these ARs, the method successfully produced sigmoids or magnetic flux ropes that were commonly observed before flares. For this reason, the method has often been used to generate initial conditions in the MHD simulations of solar flares \citep{ino15,ino18,Inoue_2021,Muhamad_2017}. Moreover, the method has also been applied for studying the magnetic field evolution of AR 12673 prior and after the X2.2 and X9.3 flares \citep{ino18,Bamba_2020,Yamasaki_2021,Inoue_2021}.

The free energy comparison was conducted only for the emergence of AR 12673 between September 4 to September 6, 2017 to evaluate the conformity between the photospheric energy calculated in this study and the volumetric energy estimated using the NLFFF model. We compared the energy evolution only for this AR because it was more complex and dynamic than other ARs. We assumed that the comparison sufficed to represent the consistency between the photospheric magnetic energy and the volumetric magnetic energy. 

\subsection{Electric Current Helicity}
Current helicity is a measure of linkage or twist of electric currents. It is often used to characterize the degree of topological complexity of a magnetic field in an AR under the assumption of force-free field \citep{Zhang_2006,Maurya_2020,Bobra_2015,Hazra_2020}. This quantity is commonly assumed to have the same sign as magnetic helicity, although recently \citet{Russell_2019} conjectured that this assumption is not true in general. Derivation of magnetic helicity from observations is impractical, so it is more convenient to derive current helicity to define the degree of AR complexity.
It is also well known that there is a hemispheric sign rule of helicity, which describes that the northern (southern) hemisphere has a dominance of negative (positive) helicity. However, previous studies showed that several flaring ARs tend to not follow the hemispheric helicity rule \citep{Pevtsov_1995, Park_2021}. In this study, we examined the helicity sign preference of our several ARNs, which were located in the southern hemisphere. For this purpose, we used SHARP metadata to obtain the derived mean current helicities of the series ARs during five CRs.

The electric current density was calculated from the given vector magnetogram data by using Ampere$'$s law, $\bf{\nabla} \times \bf{B} = \mu \bf{J}$. However, because the data only include information of a thin layer in the photosphere, only the vertical current density ($J_z$) can be calculated. The derivation using Ampere$'$s law was realized using a finite-difference method with a 9-point stencil \citep{Bobra_2014}. From the vertical current density, the proxy of the electric current helicity was estimated as follows:
\begin{equation}\label{eq6}
H_c = \bf{B_z} \cdot \bf{J_z} .
\end{equation}   

Mean current helicity for the whole AR was calculated as \citep{Bobra_2014}
\begin{equation}\label{eq67}
\overline{H_c} = \frac{1}{N}\sum \bf{B_z}\bf{J_z} .
\end{equation}   
We assumed that the mean current helicity of each AR represents the dominant current helicity at that moment. It should be noted that the error of mean current helicity given by SHARP team has included propagation errors of $B_z$ and $J_z$. We found that the errors of the current helicity were in the order of $1\times10^{-3}$ G$^{2}$/m, which was insignificant during the early period of the flaring ARs (12665 and 12673).  

\subsection{Extreme Ultraviolet Intensities}
Another alternative for obtaining the magnetic field evolution in an AR is evaluating the evolution of its extreme ultraviolet (EUV) intensities. \cite{Ugarte_Urra_2015} used the evolution of the EUV intensities of an AR observed at 304 $\mathrm{\AA}$ by the SDO/AIA to estimate the magnetic flux on that AR. They found that the peak of the EUV 304 $\mathrm{\AA}$ intensities during the AR evolution was proportional to the lifetime of that AR. Therefore, to compare the long-term evolutions of AR 12665 and AR 12673, we calculated and compared their total EUV intensities from the SDO/AIA 304 $\mathrm{\AA}$ for the AR 12665 and AR 12673. The total EUV intensities for AR 12665 were calculated from July 8, 2017 at 00 UT to July 15, 2017 at 00 UT by summing all intensities from the AR after subtracting the background intensities. The background intensities were calculated by averaging the intensities of all pixels within the center part of the solar disk. For AR 12673, the total EUV intensities were calculated from September 3, 2017 at 00 UT to September 9, 2017 at 00 UT. We avoided to utilizing the intensities data during a flare period, at least one hour before and after the flare.

\section{Results and Discussions}
\subsection{Identification of ARs}
Synoptic maps of several CRs showed that the predecessor of AR 12673 was formed long before September 2017 (see Figure 1). This was indicated by the presence of a strong magnetic field region in a similar spot to that of the AR 12673 location in CR 2194. We analyzed the AR from July 2017, when the presence of an AR was clearly observed. The locations of the ARs are marked by red circles in Figure 1. These ARs were identified as NOAA AR 12665, 12670, 12673, 12682, and 12685. Carrington coordinates of these ARNs are listed in Table 1. As listed in Table 1, the location of AR 12673 was slightly behind the estimated coordinate. This was because AR 12673 emerged in the negative (trailing) polarity region of the previous AR. Owing to the effect of differential rotation in the Sun, the AR locations were slightly shifted to the west in the Carrington maps for the latter months. 

We also tracked the active region in the EUV wavelength from July 6 to November 3, 2017, as shown in Figure 2. The results confirmed AR NOAA 12665, 12670, 12673, 12682, and 12685 as ARs reoccurring in the same region of the Sun. After selection of the best tracking rate, we found that the rotational rate of AR 12665 in 195$\mathrm{\AA}$ wavelength was approximately 26.7$^{\circ}$/day. This was consistent with the other reported results that the rotation rate of the Sun in its atmosphere observed at 193 $\mathrm{\AA}$ is slower than that in the photosphere \citep{sharma2020}. We found that AR 12665 reoccurred as AR 12670 with decaying EUV intensity over time and almost disappeared early September 2017, following which a new flux emerged as AR 12673. The EUV intensity of AR 12673 rapidly enhanced in the first three days and remained high before decaying when the AR reoccurred in the east limb as AR 12682. The AR continued to decay and became very faint when it reoccurred in the final stage as AR 12685.  

\subsection{Flaring History and ARs Properties }
Table 2 lists the period of occurrence of each identified ARN in the Earth-side solar disk and the total number of flares for each class produced by the ARN during the corresponding period. The table also lists the Hale class when each ARN formed its most complex magnetic configuration or produced flares. Clearly the ARs evolution was highly dynamic for almost five months. Only two ARNs (12665 and 12673) produced flares greater than C1-class flare, and only AR 12673 produced X-class flares. 

Table 2 also shows that there were some phase alternations between the flare productive and quiet phases. During the passage period of AR 12670, there was a pause in the flare occurrence between the periods of AR 12665 and AR 12673. Time lag in the flare occurrences of AR 12665 and 12673 suggested that the AR had potential to achieve flare periodicity of 50--55 days, which is a harmonic of solar rotation period. The periodicity was a result of the reactivation of the AR after its dormant period. This suggests that successive occurrence of flares in the subsequent rotations is not ensured. Specifically, the long-term flare periodicity of the AR, which is considered to be related to solar rotation, was most probably a consequence of the long lifetime or multiple recurrences of the AR. However, the ability of the AR to produce flares was determined by other factors during its evolution, which are discussed in the following subsection.

To examine the periodicity of the flares during the lifetimes of the ARNs, a wavelet analysis of the GOES X-ray flux time series data during 2017 was performed. Because of the interest in only long-term characteristics, the one-minute data were averaged per hour (XRF). The hourly averaged data are shown in Figure 3 (a). Because most of the flux values were very small, in the order of $10^{-8}$ -- $10^{-7}$ $W/m^{2}$, they were multiplied by $10^7$ to enhance the order of the wavelet power. Because of the significant difference in the X-ray flux between the X-class flares on September 6 and 10, 2017 and other days, these could be identified as very strong signals that almost nullified other signals during the more quiet days. Consequently, these could ignore the presence of periodicities on other days of weak X-ray flux emission from the Sun. Thus, we set a filter by applying a conditional rule to XRF(t) when it exceeds a certain limit ($v_{lim}$) as follows:
\begin{equation}
XRF(t)=  \frac{XRF(t)}{v_{lim}} + v_{lim} .
\end{equation}

Here, $v_{lim}$ was set as 50 ($W/m^2\times10^7)$. This limitation was applied because of the interest in only the pattern of the periodicity occurrence. There was little importance for the exact magnitude of the maximum signal because the extreme values ($>$ X-class flares) only occurred over three days (6, 7, and 10 September). The wavelet analysis was performed by employing a wavelet code of \cite{Torrence_1998}. This code calculates the mother Morlet wavelet of the signals that can yield the locations of the dominant frequency in the time domain. In this study, wavelet analysis was performed on the filtered XRF for each day during one year (2017). 

It can be seen from Figure 3 (b) that in only a few periods the Sun emitted strong X-ray flux. The first one was in April 2017, when several M and C-class flares occurred from AR 12644 and C-class flares from AR 12645. AR 12644 was located in the north hemisphere, whereas careful analysis using the HMI synoptic Carrington maps showed that AR 12645 was located in a region different from the region of interest, although it was also located in the southern hemisphere. The second period was associated with AR 12665 in July 2017, which released 29 C-class flares and two M-class flares, as listed in Table 2. The last one was the most prolific period when four X-class flares and many less energetic flares were released in September 2017, which originated from AR 12673. These two ARNs (AR 12665 and AR 12673) were considered to be located in the same spot, as discussed in the previous subsection. Several small pulses appeared although their magnitudes were relatively much smaller than the three periods. These small pulses might also contribute to the production of a signal of periodicity.

Figure 3 (c) and (d) show the wavelet power spectrum and the global wavelet spectrum of the X-ray flux in 2017, respectively. Clearly strong powers appeared from approximately HOY 4500 (July) to HOY 7200 (October). This spectrum also exhibits the existence of two dominant periods: approximately 25--30 days (one solar rotation) and 50--70 days (two solar rotations). Clearly, these two periodicities were related to the existence of the AR 12665 and AR 12673. The one solar rotation periodicity could have been generated from the weak X-ray flux emitted by the strong field region that periodically occurred in every rotation within this period. The two-solar-rotation periodicity was generated by the flares that originated from AR 12665 and AR 12673. 

\subsection{Magnetic Field Evolution}
Within five solar rotations, the ARs dynamically evolved, which changed its flare productivity. We investigated the magnetic field changed from the aspects of magnetic structure, total unsigned flux, and its magnetic energy for each ARN. Note that we only calculated the magnetic free energy and the total unsigned magnetic flux for the region with magnetic flux density larger than 250 G to avoid low signal to noise ratio in the weak-field region. 

\subsubsection{AR 12665}
On the first day itself since this ARN emerged from the east limb, the magnetic structure was a highly complex bipole magnetic system. The radial component of the vector magnetic field of the ARN is shown in Figure 4 (a). The bipole gradually separated farther, and many C-class and one M-class flares occurred as it approached the disk center on July 9, 2017. After these flares, the total magnetic flux and the magnetic energy decreased 20\% and 50\%, respectively, which led to the more quiet condition in the ARN. During the period from July 10 to July 13, 2017 the maximum total unsigned flux was approximately $3.1 \times 10^{22}$ Mx and the maximum of total photospheric magnetic energy was approximately $2 \times 10^{23}$ erg/cm. The evolution of this ARN after July 10 is shown in Figure 4 (b). The energy and the total magnetic flux gradually increased from July 13, 2017, following which many C-class and one M-class flares occurred again. The magnetic energy evolution in general was consistent with the total unsigned flux evolution. The recurrence of the flares in this ARN was particularly when the total magnetic flux and the photospheric magnetic energy were relatively high.

\subsubsection{AR 12670}
AR 12670 was a large bipole with a diffused magnetic flux distributed over a wide area. The radial component of the vector magnetic field of this ARN is shown in Figure 5 (a). Figure 5 (b) shows that the magnetic energy of this ARN was lower than that of the previous ARN. Moreover, both parameters presented a decreasing trend during its passage on the solar disk. Because the magnetic structure of this ARN was relatively simple, the capability of this ARN to produce a flare was low. Clearly, this ARN succeeded AR 12665 when it was decaying. 

\subsubsection{AR 12673}

This ARN was the most active ARN among the other ARNs during July to November 2017. In the beginning of its occurrence from the east limb, the ARN was very similar to AR 12670. However, on September 2, 2017 a strong magnetic flux started to emerge from the core of the ARN and rapidly evolved to create the bipole magnetic structure. The emergence of a newly emerging flux of AR 12673 and its rapid evolution to form a complex AR are shown in Figure 6. The radial component of the vector magnetic field of this ARN is shown in Figure 7 (a). Figure 7 (b) shows the continuous increase in the total unsigned flux and the photospheric magnetic energy kept increasing after the initial emergence of the strong magnetic flux. The rapid increase in the magnetic flux and the shearing and rotational motions that occurred in this region after September 2 was distinguishable from that in the previous condition. This suggests that AR 12673 was a new AR that emerged in the remnant of the old AR.

On September 4, 2017 this structure continued to evolve and already formed a very complex structure with positive and negative flux emerging in between the bipoles as well as in other parts of the core region. This complex structure survived for the next one week and produced many large flares. A detailed analysis of the photospheric magnetogram data from September 3 to 6, 2017 suggested that at least five dipoles emerged  and interacted with each other \citep{Hou_2018}. The series of magnetic dipole emergences in the north part of the main region together with the strong shearing motion and rotation created a long PIL that contained high magnetic energy accumulations in its surrounding (see Figure 8). 

Figure 8 shows how this ARN very rapidly evolved and gained its energy in one day from September 5, 2017 (Figures 8 (a) and (c)) to September 6 (Figures 8 (b) and (d)). By September 6, 2017 at approximately 11 UT, a very high magnetic free energy accumulated and reached $1.5 \times 10^{24}$ erg/cm before significantly decreasing after the X9.3 flare. There was also a decrease in the total unsigned flux after the X9.3 flare on September 6, 2017. Subsequently, the free energy tended to be constant until September 8, 2017 although the unsigned magnetic flux kept increasing. Our calculation of the magnitude and evolution of the total unsigned flux was also consistent with the total unsigned flux calculated by \cite{Moraitis2019}. Many M and X-class flares were produced even until the ARN was in the very west limb of the solar disk, including X1.3 and X8.2 flares, which occurred on September 7 and 10, 2017 respectively.

Very strong shear and rotational motion were observed from the negative patch in the east to the positive patch in the west. A flow motion caused the negative patch to penetrate its counterpart and created a strong sheared field along the PIL, which was also observed in other studies, e.g., \cite{Yang_2017} and \cite{Mitra_2018}. \cite{Bamba_2020} argued that this continuous intrusion of the negative-polarity patch into the neighboring opposite-polarity region triggered the occurrence of X2.2 and X9.3 flares on September 6, 2017 from the high free energy region near the main PIL. The free energy intensified in this region before the X9.3 flare occurred, as shown in Figure 8 (d). Using NLFFF modeling, \cite{Hou_2018} suggested that two magnetic flux ropes existed in the AR before the X9.3 flare occurred. These became unstable owing to the strong shearing motion and rotation, which caused the upper flux rope to erupt upward owing to the kink-instability. A similar scenario of multiple emergence of magnetic dipoles and the combination of the shearing motion and rotation was also believed to be responsible for the occurrence of the X8.2 flare on September 10, 2017. 

\subsubsection{AR 12682}
This ARN was quiet with no solar flares recorded during its passage on the solar disk. The magnetic structure clearly showed the decay feature of AR 12673 with some magnetic flux diffused to the wider region. The radial component of the magnetic field is shown in Figure 9 (a). Figure 9 (b) shows that the magnetic free energy decreased significantly since the previous solar rotation, and only less than $\approx$ 10\% remained. Although the total magnetic flux and the free energy increased again after October 1, 2017 there was no significant increase in the activities from this ARN. This increase was possibly due to some of the negative patches from the north entering the field of view of the SHARP image and contributing to the enhancement in the total magnetic flux density in the negative-polarity region.   

\subsubsection{AR 12685}
It is possible that AR 12682 continuously decayed to form AR 12685, which was very large in size. Consequently, this ARN appeared almost as a unipolar flux in the SHARP field of view. Figures 10 (a) and (b) show the radial component of the magnetic field in this AR and the evolutions of the total magnetic flux and the magnetic energy, respectively. This ARN was very quiet and did not show any flare potential. This AR was apparently the final stage of AR 12673, which survived for several months since September 2017. The total unsigned flux and magnetic energy of this ARN were only approximately half of those of the previous ARN. Clearly, still some small dynamic events occurred in the AR that changed the strength of its magnetic flux and its magnetic energy. For example, after 18 UT on October 27, 2017 the small positive patches in the northeast collided with each other and formed stronger large positive patches. This occurred on a smaller scale for the negative patches. This phenomenon increased both the total magnetic flux and free energy of the AR, as shown in Figure 10 (b). However, this activity did not significantly change the general trend of the magnetic evolution of the AR, which was decaying. A very weak trace of the continuation of the AR was found after AR 12685 appeared.

\subsubsection{NLFFF Magnetic Energy}
To verify the reliability and consistency of the photospheric free energy calculated using equation \ref{eq4}, it was compared with the volumetric free energy calculated using equation \ref{eq1} for the NLFFF model of AR 12673 from September 4 to September 6, 2017. AIA 171 $\mathrm{\AA}$ image of AR 12673 on September 4, 2017 at 00 UT and the corresponding  NLFFF model extrapolated by the SHARP data are shown in Figures 11 (a) and (b). The comparison of the evolution of both free energies is plotted in Figure 11 (c). 

The general trends of both free energy evolutions were consistent. Specifically, the free energies consistently increased since the emergence of the magnetic flux until the occurrence of the X2.2 and X9.3 flares. However, the volumetric free energy on the NLFFF model increased more significantly on September 4, 2017 than the photospheric free energy. This was possibly due to the presence of the magnetic flux rope in the NLFFF model, which could not be observed from the photospheric magnetogram data. The magnetic flux rope was also reported to form in many other studies on the NLFFF model of AR 12673 \citep{Hou_2018,ino18,Liu_2018,Zou_2020,Yamasaki_2021} or in data-driven simulation \citep{Price_2019}. The magnetic free energy of the NLFFF drastically decreased after the X9.3 flare on September 6, 2017. As has been described earlier, the photospheric free energy also decreased after this flare occurred. 
   
It is important to note again that equation \ref{eq4} calculates the nonpotentiality of the AR, and not the real free energy. From the comparison shown in Figure 11 (c), it can be inferred that the evolution of the photospheric free energy calculated using equation \ref{eq4} was in general consistent with that of the 3-D magnetic free energy calculated using the NLFFF model. The NLFFF model provided more information about the 3-D structure of the magnetic field in the corona, which contributed to the more realistic estimation of the free energy magnitude. However, this comparison confirms that the equation \ref{eq4} yields quite reliable results for representing the free energy evolution of the AR.   

\subsection{Active Longitude and Current Helicity}
We have shown that AR 12665, 12670, 12673, 12682, and 12685 occurred in the same region of the Sun. AR 12670 was clearly the continuation of AR 12665 that reoccurred in the following disk passage. After one Carrington rotation, AR 12673 emerged in the extended plage of the negative (trailing) polarity of the decayed AR 12670 (see Figure 6). AR 12673 grew rapidly and formed a very complex magnetic structure in just a few days. After one solar rotation, this AR became AR 12682 that was less complex than its predecessor. In its final stage, the AR reoccurred as AR 12685 before it totally decayed and disappeared.

The occurrences of the subsequent ARs over multiple rotations in the same region could indicate the presence of an active longitude. We found that the Carrington longitudes of the ARNs during 5 CRs spanned between 111$^{\circ}$-133$^{\circ}$. This narrow longitude band was close to the sunspot active zone near 90$^{\circ}$ of longitude that was commonly observed over multi solar rotations \citep{Ivanov_2007}. Active longitude near 90$^{\circ}$ of Carrington longitude in the southern hemisphere was also observed in the first Carrington rotations in solar cycle 24 \citep{Komm_2015}. However, we did not find any AR in the southern hemisphere that has longitude separation approximately 180 degrees with the studied ARs during 5 CRs. The opposite active longitude was often observed to form a pair of active longitude in the Sun in the past solar cycles \citep{Usoskin_2005, Mandal_2017}. However, we found two ARs (AR 12664 and 12671) that occurred in the northern hemisphere that were separated approximately 180 degrees of longitude from the studied ARs in the time period from July to November 2017. Among these two ARs, only AR 12671 that produced many flares.

The magnetic field characteristics of AR 12673 are distinctive compared to other ARNs, particularly in the amount of total magnetic flux, magnetic energy, flux emergence rate, and rotational and translational motions. The rapid emergence of strong magnetic flux in the pre-existing AR may be related to the prolific flare-productivity of AR 12673.
Previous studies suggested that large-scale converging \citep{Gizon_2001} and vortex \citep{Komm_2007} flows occurred around active regions. The combinations of the two flows are believed to be more effective in an AR near the active longitudes because many ARs frequently occurred  there. It indicates that some turbulence motions may play a role to rise flux tubes from the subsurface and disturb the hemispheric sign preference (HSP) of an AR \citep{Komm_2015}. Moreover, the presence of such local turbulence below a pre-existing AR may contribute to generate new emerging fluxes that eventually create a very strong and complex AR that potentially produces large flares \citep{Komm_2015,Park_2021}. Our study shows that the emergence of AR 12673 in the pre-existing AR (AR 12670) had some characteristics that were in agreement with this scenario. This could give a potential explanation for the flare-productive properties of AR 12673.

The peculiarities of the AR 12673 has been qualitatively discussed by \citet{Getling_2019}, who found that the emergence of the new emerging flux in this AR was similar to the pattern of a fluid flow in a roundish body. He suggested that the dynamics of the sunspot in this AR was related to the surface layer that emerged above the old cluster that was rooted to the deeper layers of the convection zone.
           
Figure 12 shows mean current helicities of 5 ARNs during its passage in the solar disk calculated using equation \ref{eq67}. It can be seen that the current helicities of the flaring ARNs (AR 12665 and 12673) were relatively high during their first rotations. The helicities decreased when the ARs rotated for the second and the third times. During their early occurrences, ARs have negative helicities. However, at the last rotations, the helicities were very low and the amount of helicities became comparable to the error bars. During these last periods of the ARs, the signs of helicities were mix between positive and negative. The decreasing helicity indicated that the magnetic helicity has been largely transferred from the photoshere to the corona or even to the interplanetary space through some eruptions during the lifetime of the ARs.

We showed that AR 12665, 12673, 12682 have negative sign of current helicities. The negative current helicity sign of AR 12673 was also observed in other studies, e.g., \citep{Yan_2018,Vemareddy_2019,Moraitis2019}. This sign was anti-HSP for the southern hemisphere. The heliographic region around the Carrington longitude of the ARs in our study, which was anti-HSP, had been found to be more flare productive than other heliographic regions \citep{Park_2021}. It was well known that active regions that produced many flares tended to not follow HSP rule \citep{Pevtsov_1995, Maurya_2020}. Active regions that have strong magnetic fields are also found to be anti-HSP \citep{Zhang_2006}. This tendency was consistent with the mean-field dynamo theory that the opposite helicity signs are produced in the mean field and small-scale fluctuations \citep{Zhang_2006,Komm_2015}. Our study showed that the ARs that occurred during July-October 2017 confirmed previous findings that the strong ARs and flare-productive ARs tended to oppose HSP.

\subsection{EUV Intensities Evolution}
           
EUV intensity evolutions of AR 12665, 12670, 12673, 12682, and 12685 were studied to understand their relation with the magnetic field evolution of the ARNs. It was believed that EUV intensity evolution of an AR was linked to the evolution of its magnetic field \citep{Ugarte_Urra_2015}. The total EUV intensity of an AR was also found to be closely related to the area of AR \citep{Verbeeck_2013}.

Figure 13 shows the comparison of the light curves of the EUV 304 $\mathrm{\AA}$ for AR 12665 and AR 12673. The intensities of AR 12673 were higher than those of AR 12665. This is consistent with the higher total magnetic flux amount of AR 12673 (see Figures 4 and 7). \cite{Ugarte_Urra_2015} found that the peak intensity of an AR observed at 304 $\mathrm{\AA}$ was also proportional to its lifetime. Because the peak intensity of AR 12673 was approximately thrice higher than that of AR 12665, the lifetime of AR 12673 was expected to be longer than that of AR 12665, which was actually found. However, the lifetime of AR 12665 was not one-third of that of AR 12673. Note that AR 12665 had not really disappeared when AR 12673 emerged, which made the precise comparison of their lifetime difficult.

In general, we found that the EUV intensities of the ARs were proportional to the total magnetic flux of the ARs. This means that the stronger magnetic field of an AR, the brighter the AR in the EUV wavelength. Nevertheless, the peak EUV intensities did not always have constant proportion with the total magnetic flux. For example, AR 12673, which was three times brighter than AR 12665, has a total magnetic flux that was two times stronger than that of AR 12665.

Evolution of the EUV intensities also has the same trend with the total magnetic flux evolution. It can be seen from Figure 13 that the EUV intensity of the flaring AR 12665 and 12673 were decreasing until they disappeared in the third disk passage. Therefore, we found that EUV intensity evolution of the AR could be a proxy for the total magnetic flux evolution. This is in agreement with other results that studied the long-term evolution of EUV intensities of several ARs in solar cycle 24 \citep{Ugarte_Urra_2015,IGLESIAS20201641}. However, we could not find any clear signature of a flare precursor in the EUV intensities evolution of the flaring ARs prior to the flares. The only indication of the occurrence of the flaring AR was a rapid enhancement of the EUV intensity during the early occurrence of the newly emerging flux of AR 12673 (see Figure 13). This rapid increase of EUV intensities infers that the dynamics of the coronal magnetic field is closely linked with the dynamics of the photospheric magnetic field.

\section{Summary and Conclusions}
We showed and identified the recurrent ARs from July to November 2017 over multiple solar rotations. Although there was a differential rotation effect, the traces of the ARs were still clearly distinguishable. The magnetic field of each AR dynamically evolved periodically and affected its productivity. The magnetic field structures of the ARs and their properties significantly differed over multiple rotations. 

We found that although AR 12673 apparently occurred in the same region as AR 12665 and AR 12670, this ARN was actually a result of a new emerging flux on a pre-existing fading AR. This suggested that AR 12673 was a distinguished magnetic system that was different from its predecessors, which was AR 12665. The main differences between these two ARs were the emerging and the complexity growth rates. These were clearly observed from the variations in the development rate and magnitude of the photospheric free energy evolution. AR 12673 had a much higher emerging rate and became a very complex AR owing to the combination of the shearing and rotating motions. A difference was also observed in the intensities of EUV 304 $\mathrm{\AA}$ at the peak times of AR 12665 and AR 12673. Specifically, AR 12673 produced almost three times stronger EUV 304 $\mathrm{\AA}$ intensities than its predecessor.  

AR 12665 and 12673 occurred in the magnetic activity complex that was located in the southern hemisphere spanned from 111$^{\circ}$--133$^{\circ}$ of Carrington longitude. This location was suspected to be associated with the active longitude near 90$^{\circ}$ of Carrington longitude that has been long observed in the southern hemisphere for multiple solar cycles. The emergence of AR 12673 in the pre-existing AR 12670 indicated that the subsurface magnetic activity near the active longitude played an important role to generate strong magnetic flux with high complexity that led to the occurrence of the flare-productive AR. \citet{Komm_2015} suggested that the combination of sub-photospheric converging and vortex flows in a magnetic complex can generate a newly emerging flux that  interacts with pre-existing AR to create a more complex AR. \citet{Getling_2019} found that the emergence of magentic flux of AR 12673 in the pre-existing AR 12670 was similar to a fluid flow in a roundish body.

The two flaring ARs (12665 and 12673) had negative current helicities, which were opposite to the helicity sign preference in the southern hemisphere. In the subsequent rotations, helicity signs of AR 12665 (as AR 12670) and AR 12673 (as AR 12682 and then 12685) tended to be in agreement with the HSP. \citet{Komm_2015} suggested that the dynamics of magnetic field in the convection zone near the active longitude was responsible for the creation of AR that opposed HSP. Anti-HSP AR was commonly found to emerge in the pre-existing magnetic flux of prior AR \citep{Komm_2015}. It is possible that the emergence of the newly emerging flux of AR 12673 in the remnant of the prior AR (12670) was related to the abovementioned process. This could explain why the AR 12673 had a negative helicity sign and produced so many flares. By analyzing relative magnetic helicities in the Sun for almost one solar cycle (solar cycle 24), \citet{Park_2021} also found that the heliographic region where AR 12673 occurred was very flare-productive and strongly dominated by negative helicity.

Our results show that the build-up of the nonpotentiality in AR 12673 was rapid in the early stage of its emergence, and it soon became flare-productive. Subsequently, the AR decayed and did not produce any flare during its long decaying stage. This behaviour was consistent with previous results of \cite{1999ASPC..184..302V} and \cite{vdg2003} who studied the long-term evolution of AR 7978. We found that the AR became flare-productive when a strong magnetic flux emerged and formed 	a relatively complex structure within 3--5 days. The complex structure could have been created by the shearing and rotating motions in the photosphere that simultaneously occurred during the emergence of the strong magnetic flux. These mechanisms--the emergence of strong magnetic flux and the shearing or rotating motions in the photosphere--were the main determinants of the flare productivity of the long-lived AR (see \cite{Toriumi2019} for the characteristics of flare-productive ARs). 

Because AR 12665 and AR 12673 emerged at the same location, they could be sources for flare periodicity in a multi-rotations time range. In the ARs considered in this study, flares occurred only in two ARNs that were separated by one solar rotation. This might be interpreted as an indicator of the harmonics of the solar rotation in the flare periodicity. Furthermore, the existence of the flare periodicities during July--November 2017, which are related to solar rotation and its harmonics, was due to the long lifetime and dynamics of the AR. This long-term flare periodicity itself cannot be used as an early warning for solar flares unless an AR maintains its magnetic structures for more than one rotation. However, the magnetic structures of the recurrent ARs in our study changed significantly over multiple rotations. This suggests that the complexity of the AR did not survive for more than one rotation. The reactivation of the AR, most probably requires a new strong emerging flux that emerges on the decaying AR.

We revealed that the properties of the reoccurring ARs and their relations with the location of active longitude might be useful for a long-term prediction of solar flares. Our results show that it is important to conduct long-term observations of ARs to identify the locations of active longitudes in order to anticipate the occurrence of a flaring AR. However, the timescale of the magnetic energy build-up in the flaring AR was on the order of several days. Flares tended to occur when the total unsigned flux and the total photospheric free magnetic energy were relatively high. Both parameters could be fluctuated by the flux emergence rate and the dynamics motion in the photosphere within a time range of 2--3 days. Therefore, advancing flare prediction by more than one week is still very challenging with only the information obtained from the magnetogram data and even flare history.   

\normalem

\begin{acknowledgments}
We thank the anonymous reviewer whose comments and suggestions helped improve and clarify this manuscript. This research is supported by Space Research Center, LAPAN/BRIN. HMI is an instrument onboard the SDO, a mission for NASA$'$s Living with a Star Program. We thank the SHARP team for providing and maintaining the magnetogram data used in this study. Python wavelet software was provided by Evgeniya Predybaylo based on \cite{Torrence_1998} and is available at URL: \url{http://atoc.colorado.edu/research/wavelets/}. NLFFF code was run in the super computer of Nagoya University, Japan. Imagery of NLFFF is produced by VAPOR \citep{Li_2019}.

\end{acknowledgments}

\bibliography{bibtex}{}
\bibliographystyle{aasjournal}

\newpage
\begin{deluxetable*}{cccccc}
\tablenum{1}
\tablecaption{Carrington coordinates of the Active Regions}
\tablewidth{0pt}
\tablehead{
\colhead{ARN} & \colhead{CR} & \colhead{Date} & \colhead{Carr.lon$^{a}$} &
\colhead{Carr.lon$^{b}$} & \colhead{Carr.lat$^{b}$} \\
}
\decimalcolnumbers
\startdata
12665 & 2192 & 2017-07-12 00:00 UT & 111 & 111 & -6 \\
12670 & 2193 & 2017-08-08 00:00 UT & 118.6 & 121 & -6 \\
12673 & 2194 & 2017-09-04 00:00 UT & 124.7 & 117 & -10 \\
12682 & 2195 & 2017-10-01 00:00 UT & 129 & 124 & -11  \\
12685 & 2196 & 2017-10-27 00:00 UT & 132.7 & 132 & -9 \\
\enddata
\tablecomments{\begin{small}
$^{a}$ Calculated by equation \ref{eq01} at the corresponding time in the date column.\\
$^{b}$ Coordinates given by the Heliophysics Event Catalogue at the corresponding time in the date column.
\end{small}
}
\end{deluxetable*}

\begin{deluxetable*}{ccccccc}
\tablenum{2}
\tablecaption{Number of Flares and Properties of the Active Regions}
\tablewidth{0pt}
\tablehead{
\colhead{ARN} & \colhead{Period} & \colhead{C-class} & \colhead{M-class} &
\colhead{X-class} & \colhead{Hale Class (max)} \\
}
\decimalcolnumbers
\startdata
12665 & July 5--18, 2017 & 29 & 2 & - & $\beta\gamma$  \\
12670 & Aug 1--14, 2017 & - & - & - & $\beta$ \\
12673 & Aug 28 -- Sep 10, 2017 & 55 & 27 & 4 & $\beta\gamma\delta$  \\
12682 & Sep 25 -- Oct 7, 2017 & - & - & -  & $\beta$  \\
12685 & Oct 20 -- Nov 2, 2017 & - & - & - & $\alpha$  \\
\enddata
\end{deluxetable*}

\newpage
\includegraphics[scale=0.41]{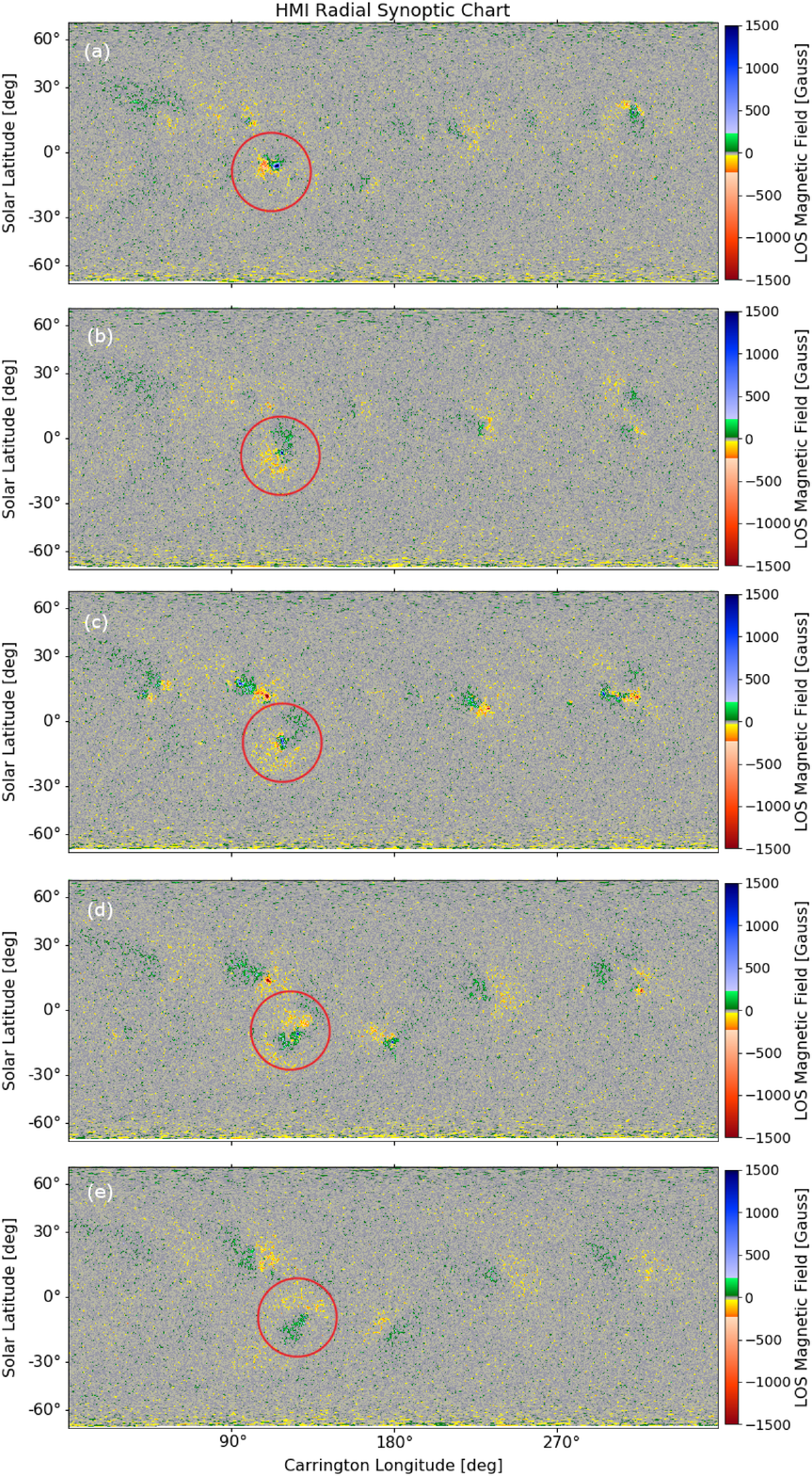}\label{fig1}
\\
Fig.\ 1 (a)--(e) Carrington maps of HMI radial magnetic field for the CRs 2192--2196, respectively. Horizontal axis represents Carrington longitude, and vertical axis shows sine latitude. Red circles show locations of corresponding AR during five CRs. 
\\

\includegraphics[scale=0.35]{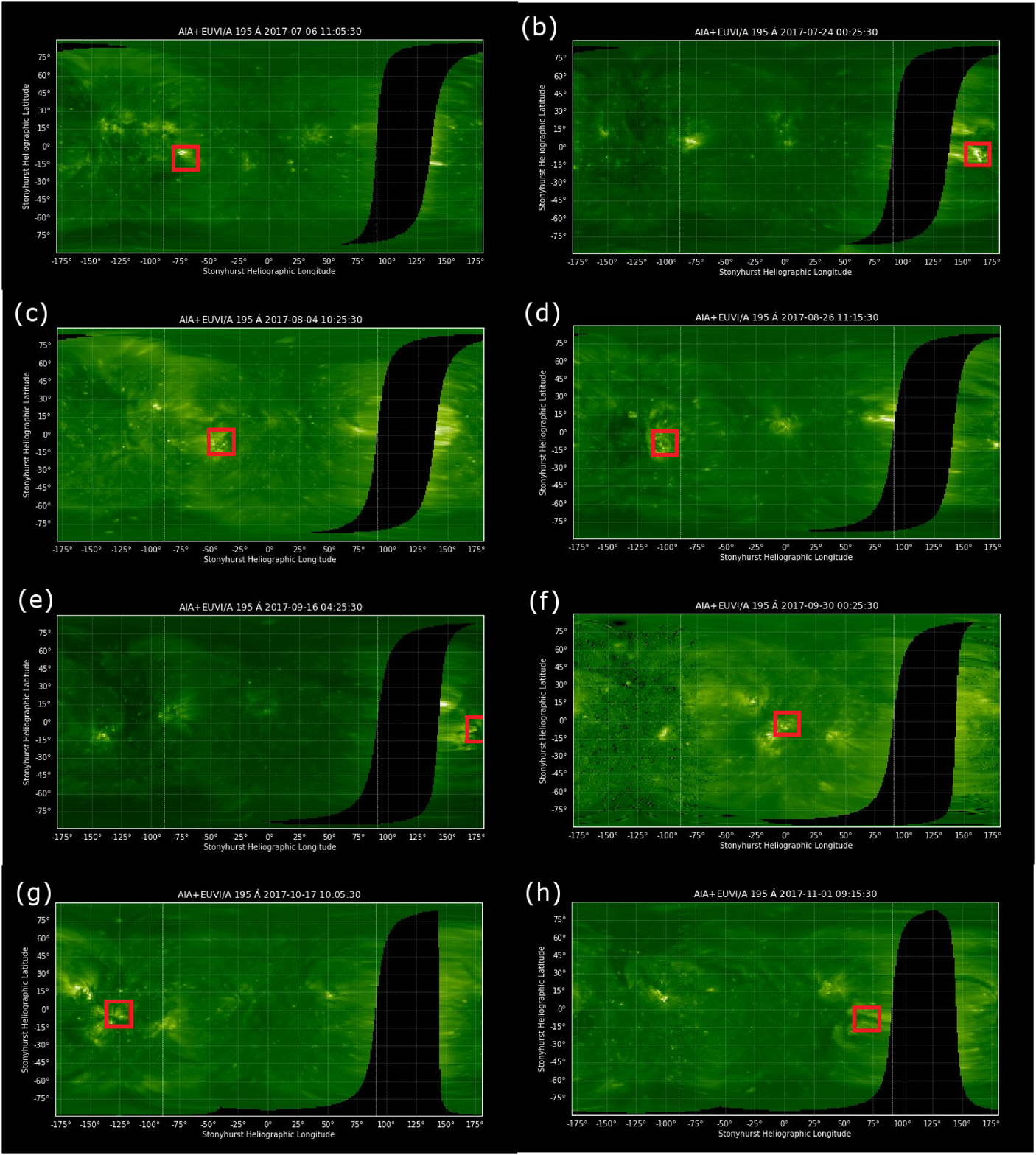}\label{fig2}
Fig.\ 2 (a) Fullmaps of Sun from AIA 193 $\mathrm{\AA}$ and EUVI-A 195 $\mathrm{\AA}$ on July 6, (b) July 24, (c) August 4, (d) August 26, (e) September 16, (f) September 30, (g) October 17, (h) and November 1, 2017. Red squares mark locations of tracked AR.   
\\
\\

\includegraphics[scale=0.35]{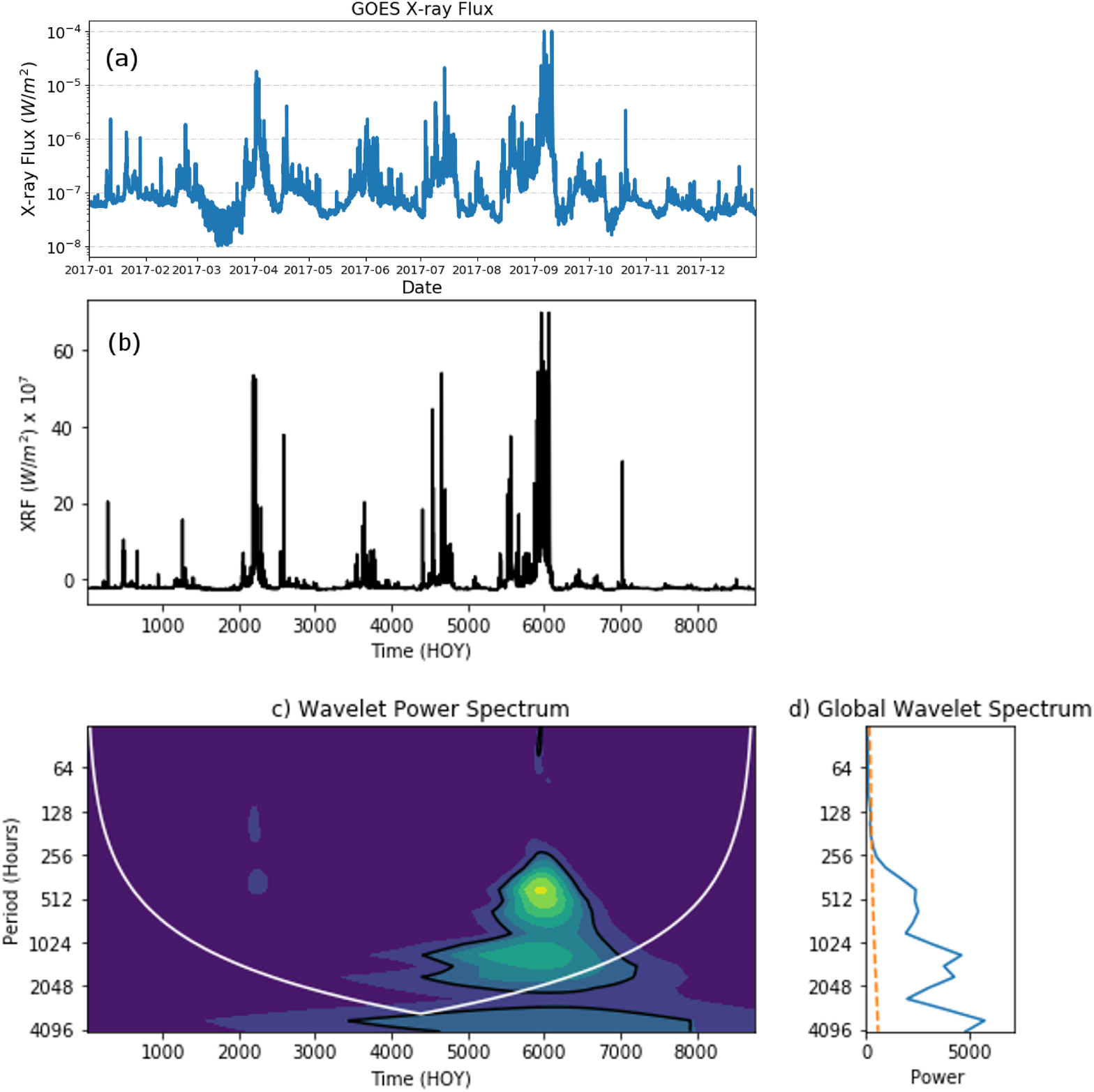}\label{fig3}

Fig.\ 3 (a) Hourly averaged GOES X-ray flux from January 1 to December 31, 2017. (b) Filtered and multiplied GOES X-ray flux. X-axis is in hours of year (HOY) unit. (c) Wavelet power spectrum and (d) global wavelet plot of GOES X-ray flux in 2017. White line is cone of influence below which area is strongly affected by edge effect. Black contours in wavelet plot and red dotted line in global wavelet plot represent 95 $\%$ of significance level.
\\
\\

\includegraphics[scale=0.8]{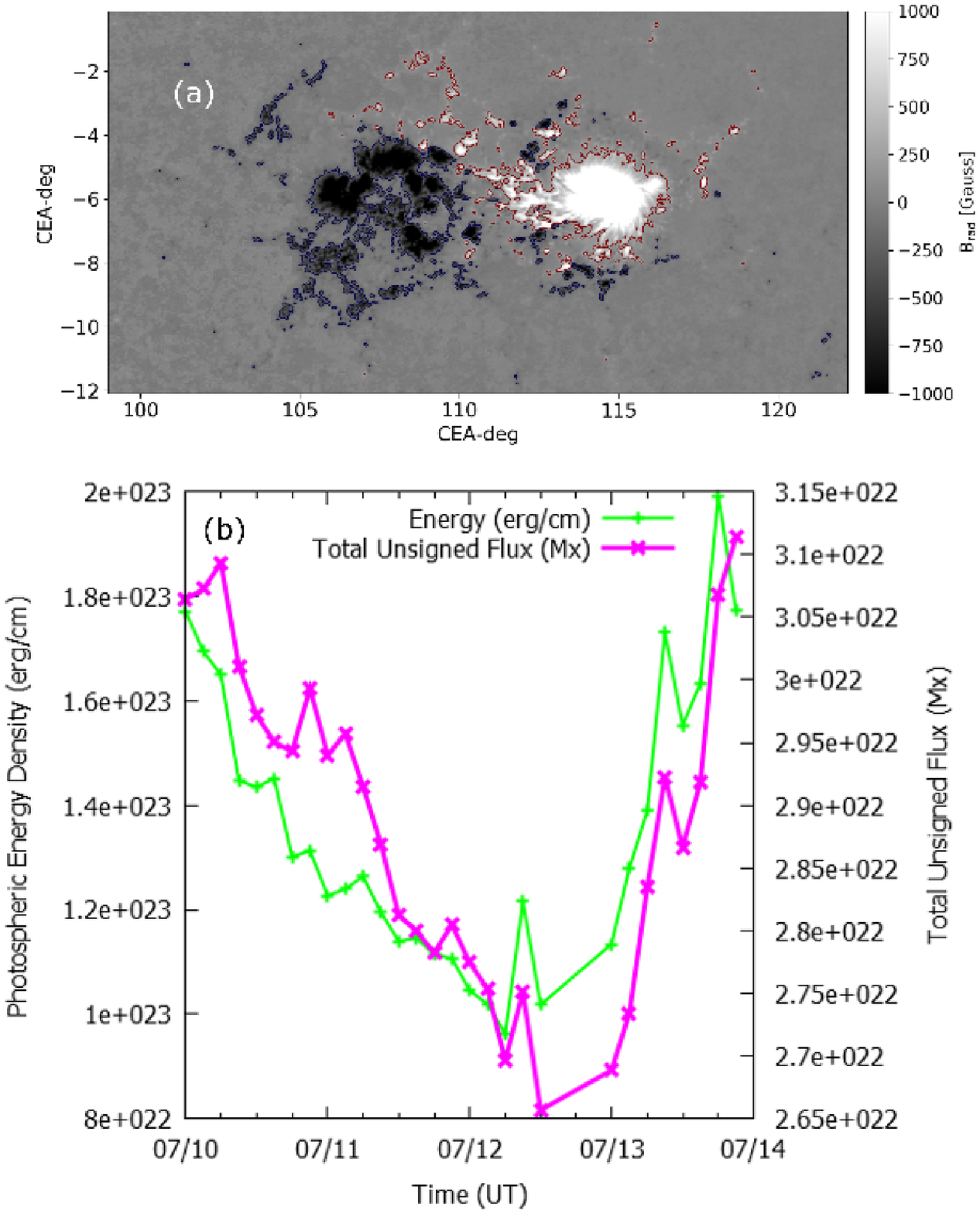}

Fig.\ 4 (a) Radial component of vector magnetic field of AR 12665 on July 10, 2017 at 03:00 UT. Red and blue contours represent magnetic flux densities of 250 and -250 G, respectively. (b) Time evolutions of total unsigned flux and total photospheric magnetic free energy corresponding to this AR. 
\\

\includegraphics[scale=0.8]{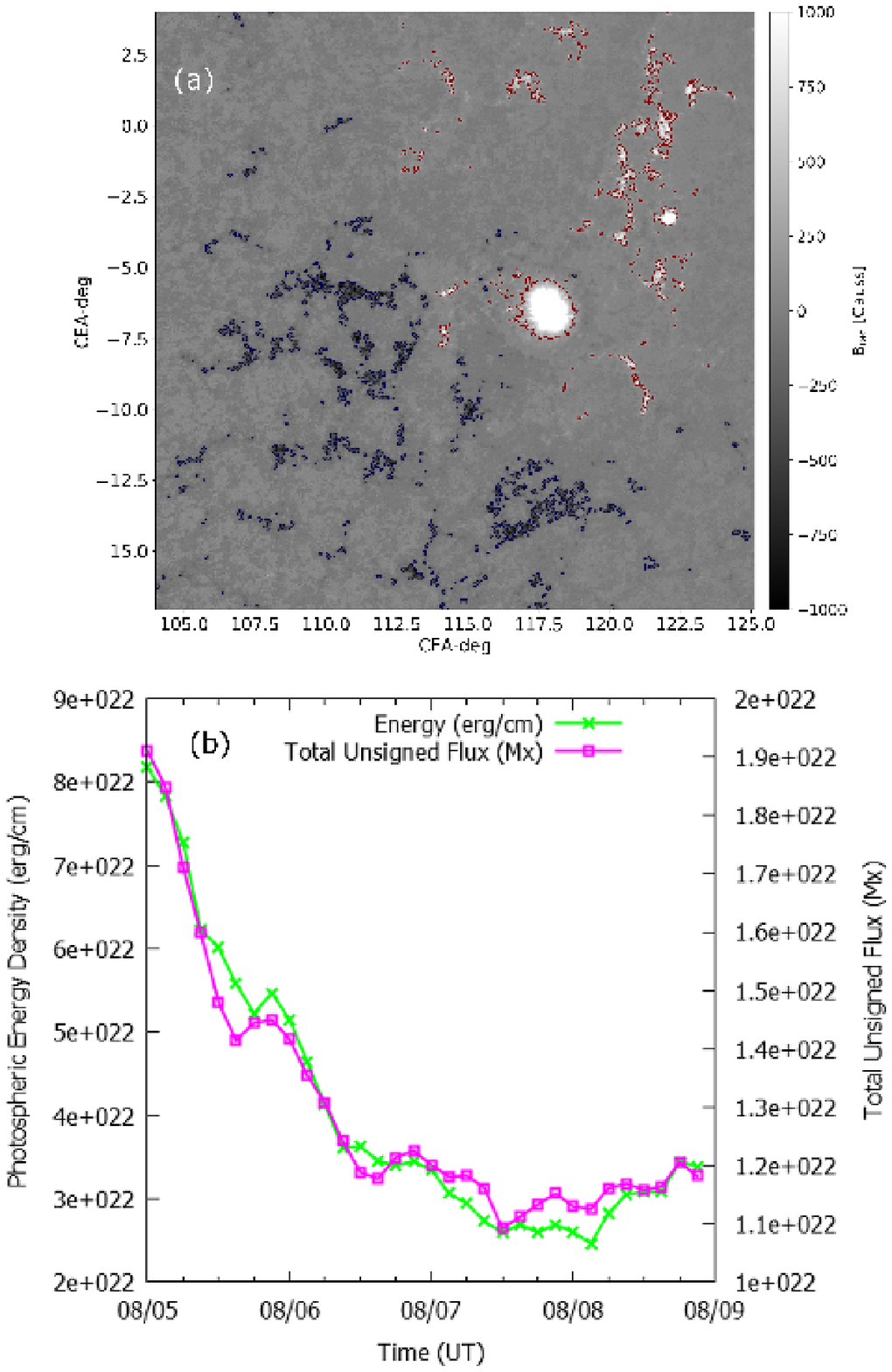}

Fig.\ 5 (a) Radial component of vector magnetic field of AR 12670 on August 5, 2017 at 12:00 UT. Red and blue contours represent magnetic flux densities of 250 and -250 G, respectively. (b) Time evolutions of total unsigned flux and total photospheric magnetic free energy corresponding to this AR. 
\\

\includegraphics[scale=0.5]{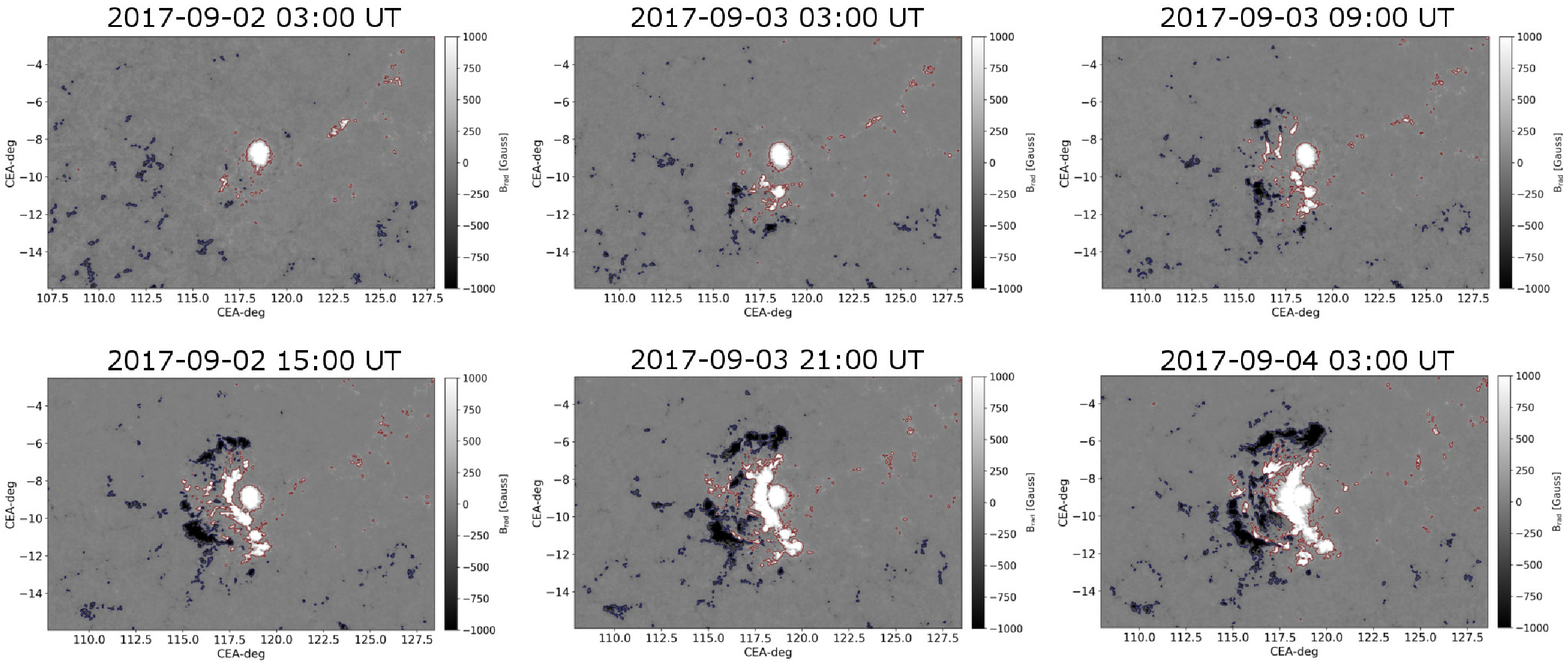}
\\
Fig.\ 6 Evolution of newly emerging flux of AR 12673 in negative (trailing) polarity of decaying AR 12670.

\includegraphics[scale=0.8]{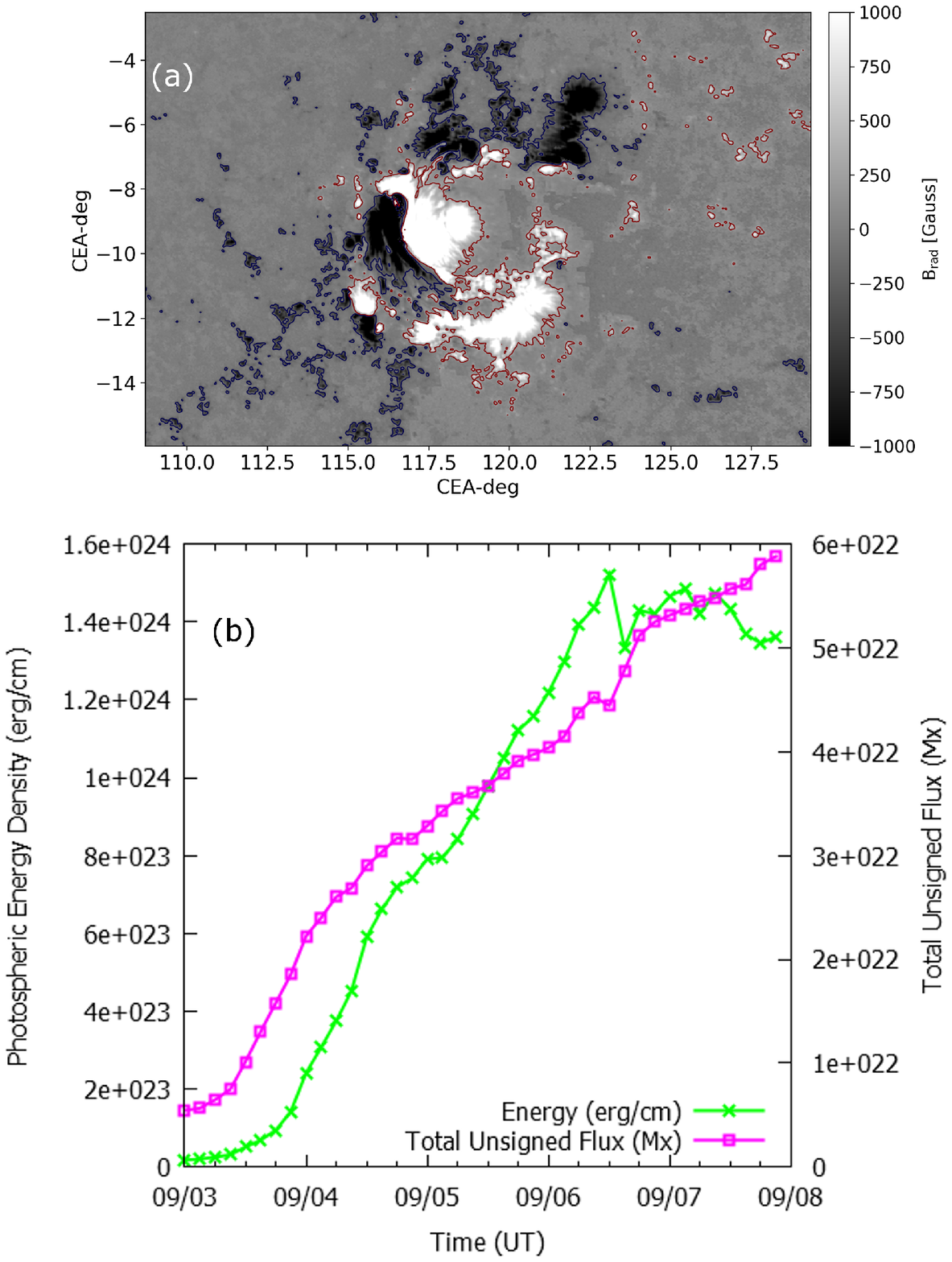}

Fig.\ 7 (a) Radial component of vector magnetic field of AR 12673 on September 6, 2017 at 06:00 UT. Red and blue contours represent magnetic flux densities of 250 and -250 G, respectively. (b) Time evolutions of total unsigned flux and total photospheric free magnetic energy corresponding to this AR. 
\\
\\

\includegraphics[scale=0.22]{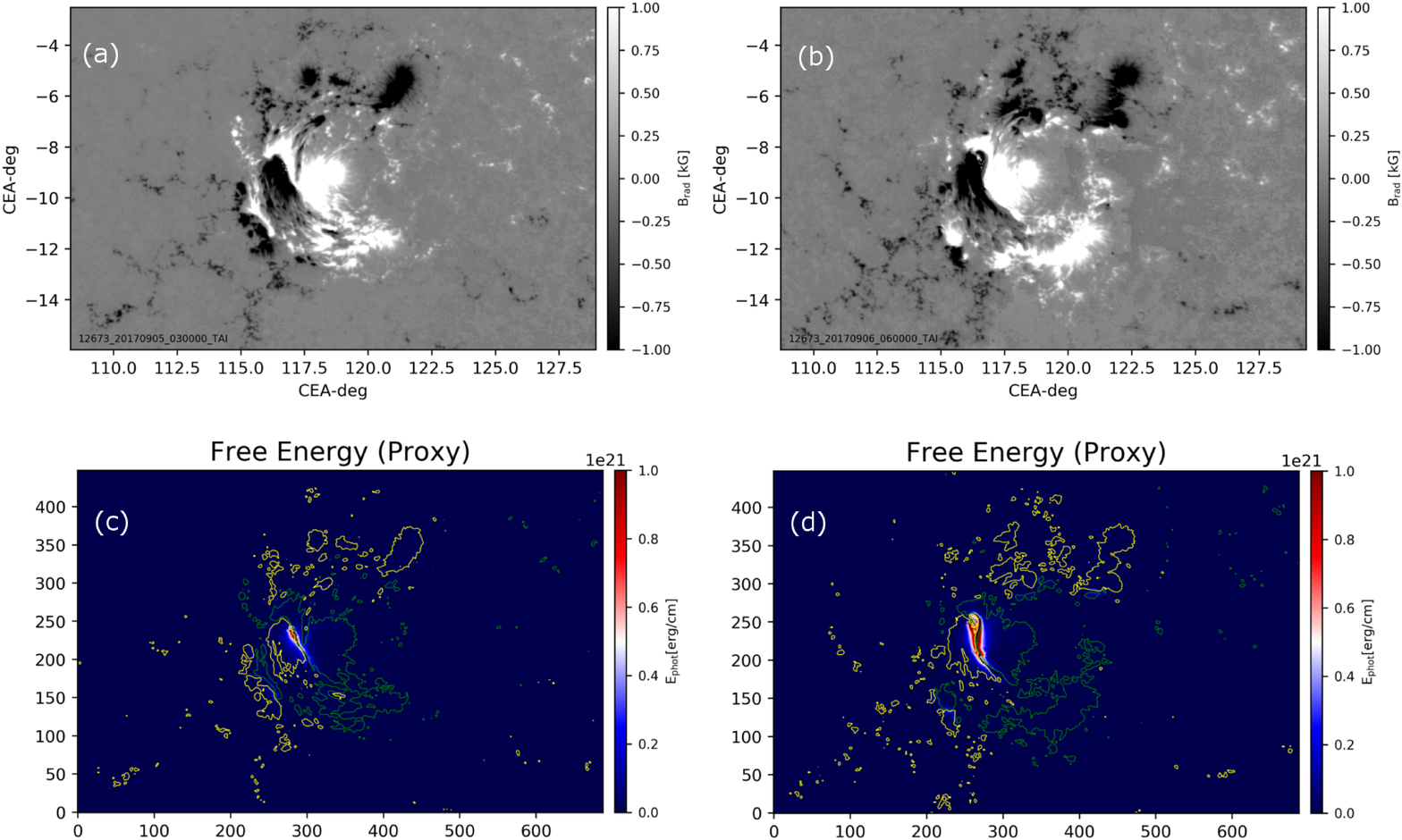}

Fig.\ 8 (a) Radial component of magnetic field of AR 12673 at 03:00 UT on September 5, 2017 and (b) 06:00 UT on September 6, 2017. (c) Distribution of photospheric magnetic energy of AR 12673 at 03:00 UT on September 5, 2017 and (d) 06:00 UT on September 6, 2017. Green and yellow contours represent magnetic flux densities of 500 and -500 G, respectively.
\\
\\

\includegraphics[scale=0.8]{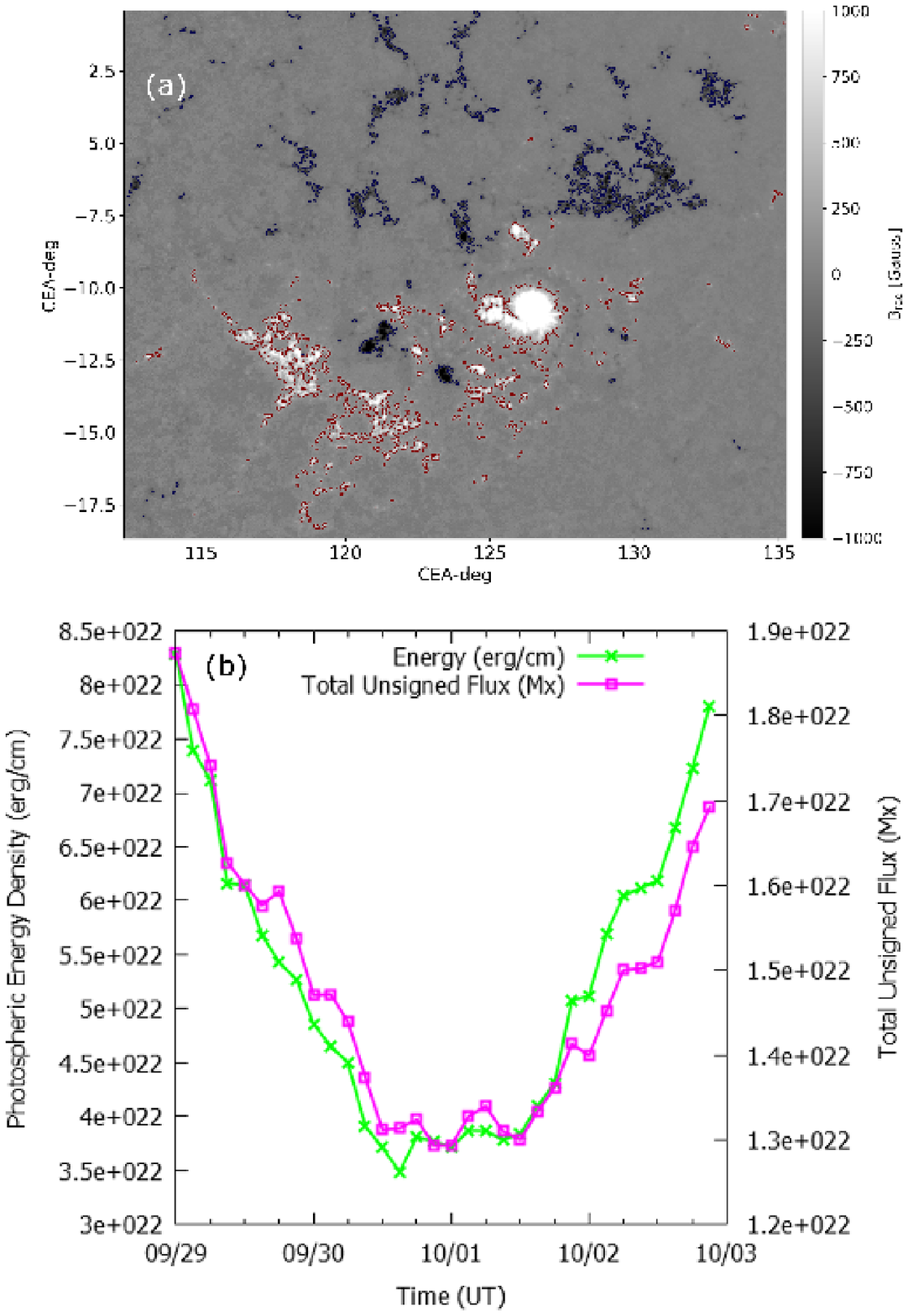}

Fig.\ 9 (a) Radial component of vector magnetic field of AR 12682 on September 29, 2017 at 03:00 UT. Red and blue contours represent magnetic flux densities of 250 and -250 G, respectively. (b) Time evolution of total unsigned flux and total photospheric magnetic free energy corresponding to this AR. 
\\

\includegraphics[scale=0.8]{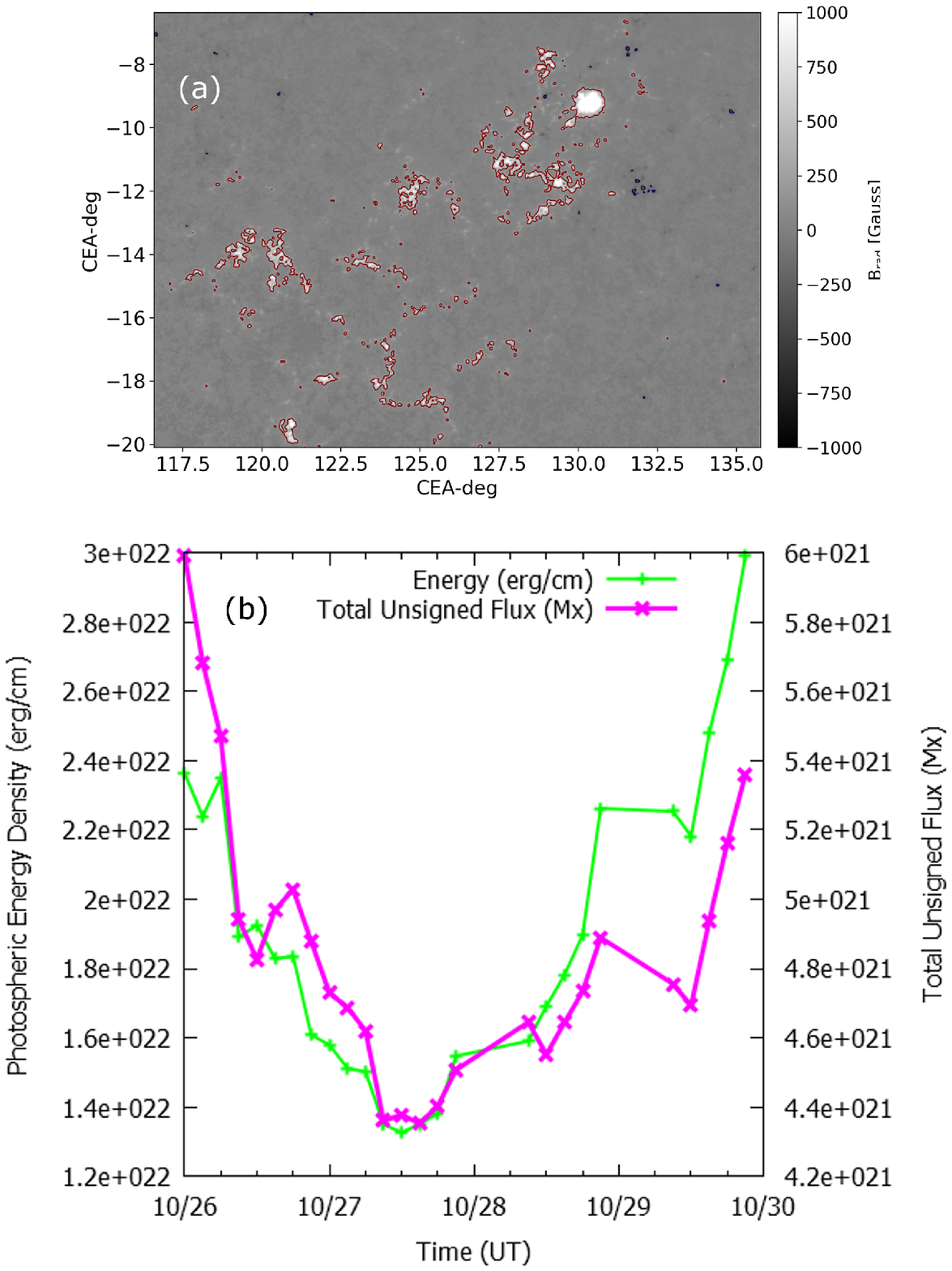}

Fig.\ 10 (a) Radial component of vector magnetic field of AR 12685 on October 27, 2017 at 03:00 UT. Red and blue contours represent magnetic flux densities of 250 and -250 G, respectively. (b) Time evolution of total unsigned flux and total photospheric magnetic free energy corresponding to this AR. 
\\

\includegraphics[scale=0.4]{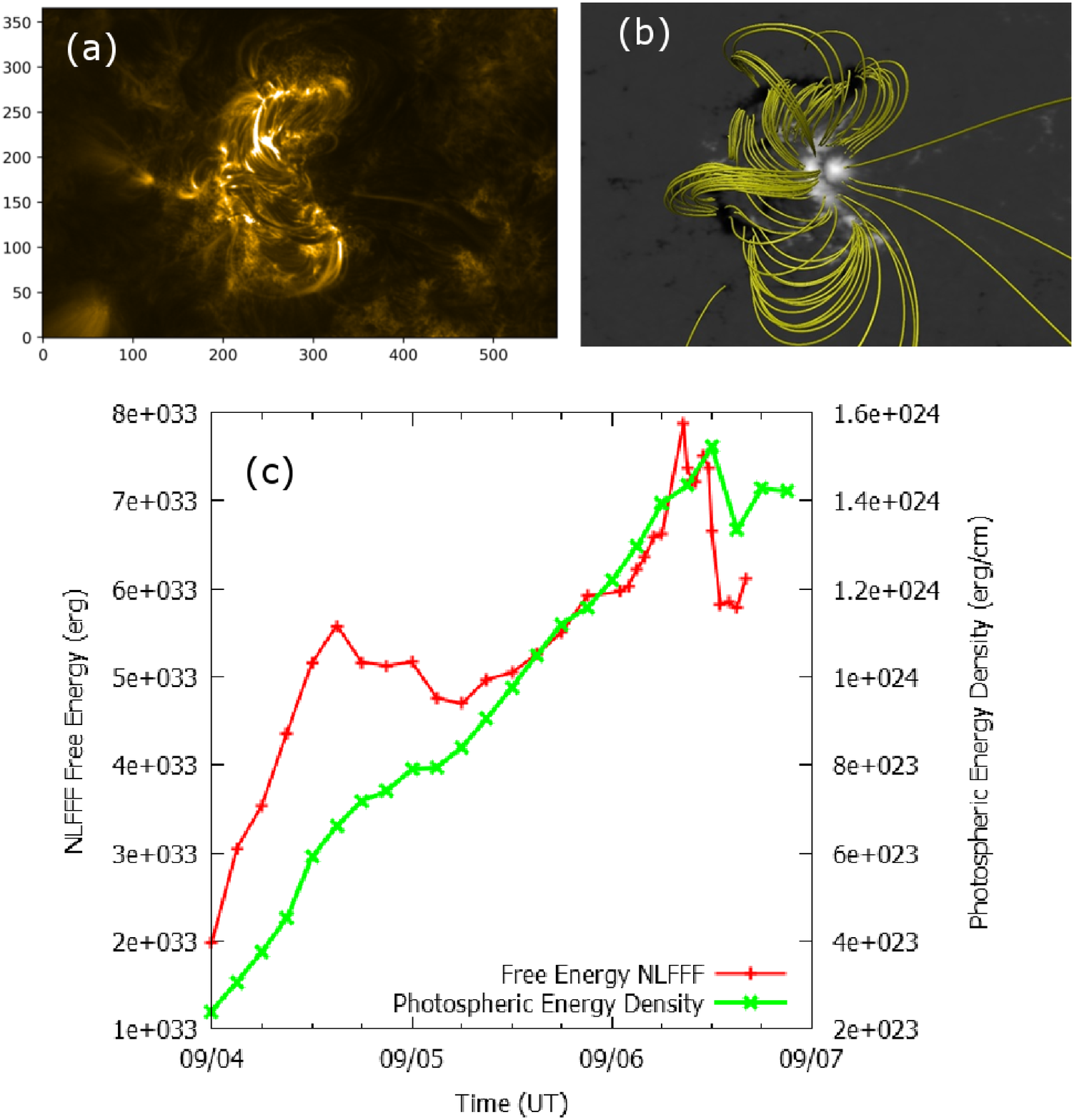}
\\
Fig.\ 11 (a) AR 12673 on September 4, 2017 at 00:00 UT observed by AIA 171 $\mathrm{\AA}$ instrument. (b) NLFFF model of AR 12673 for bottom boundary condition given by SHARP data at same time as in (a). (c) Time evolutions of total NLFFF magnetic free energy and total photospheric free energy of AR 12673. 
\\

\includegraphics[scale=0.2]{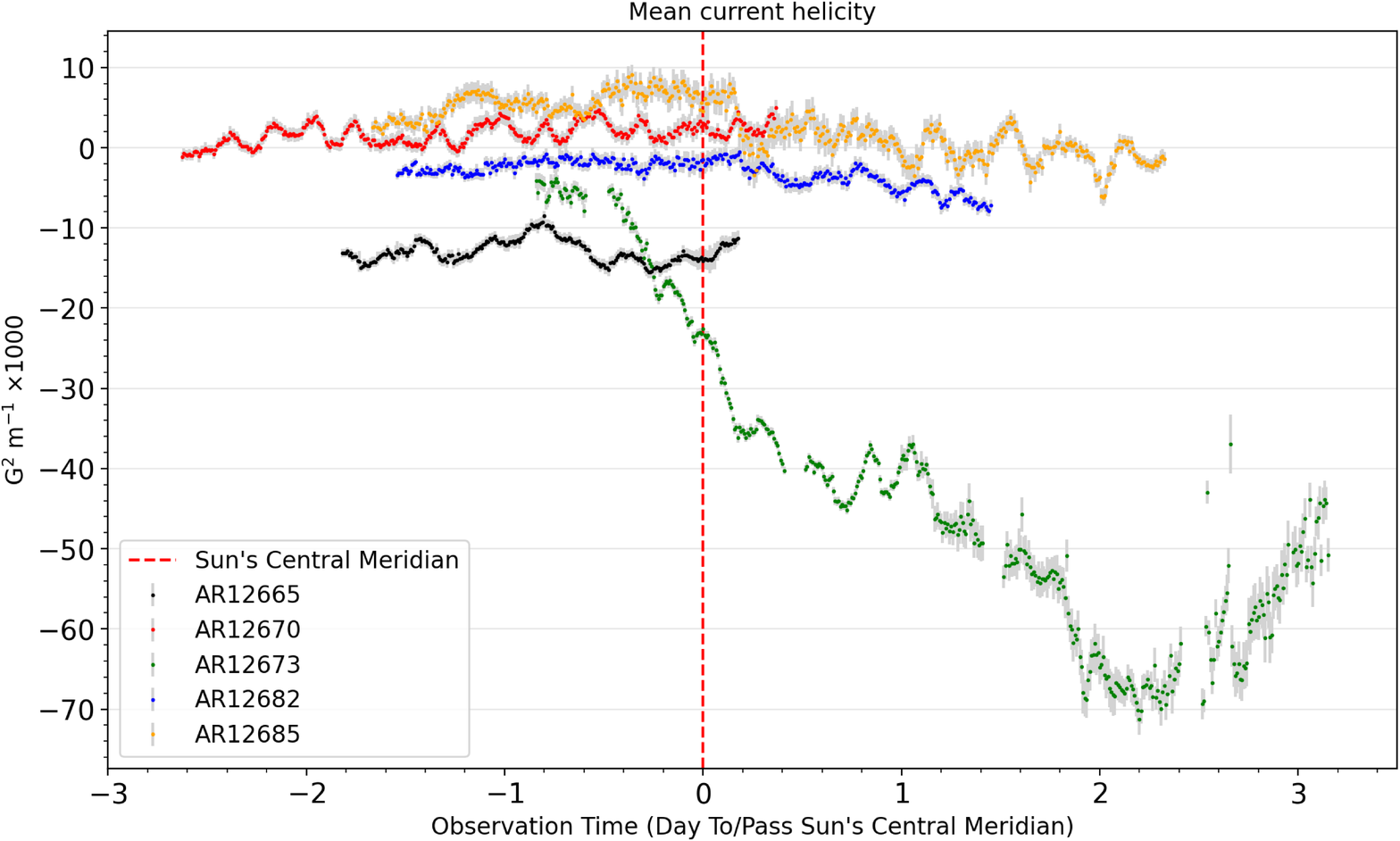}
\\
Fig.\ 12 Current helicities of AR 12665, 12670, 12673, 12682, and 12685 during its transit near central meridian. Light gray lines represent error bars of current helicities. 
\\

\includegraphics[scale=0.65]{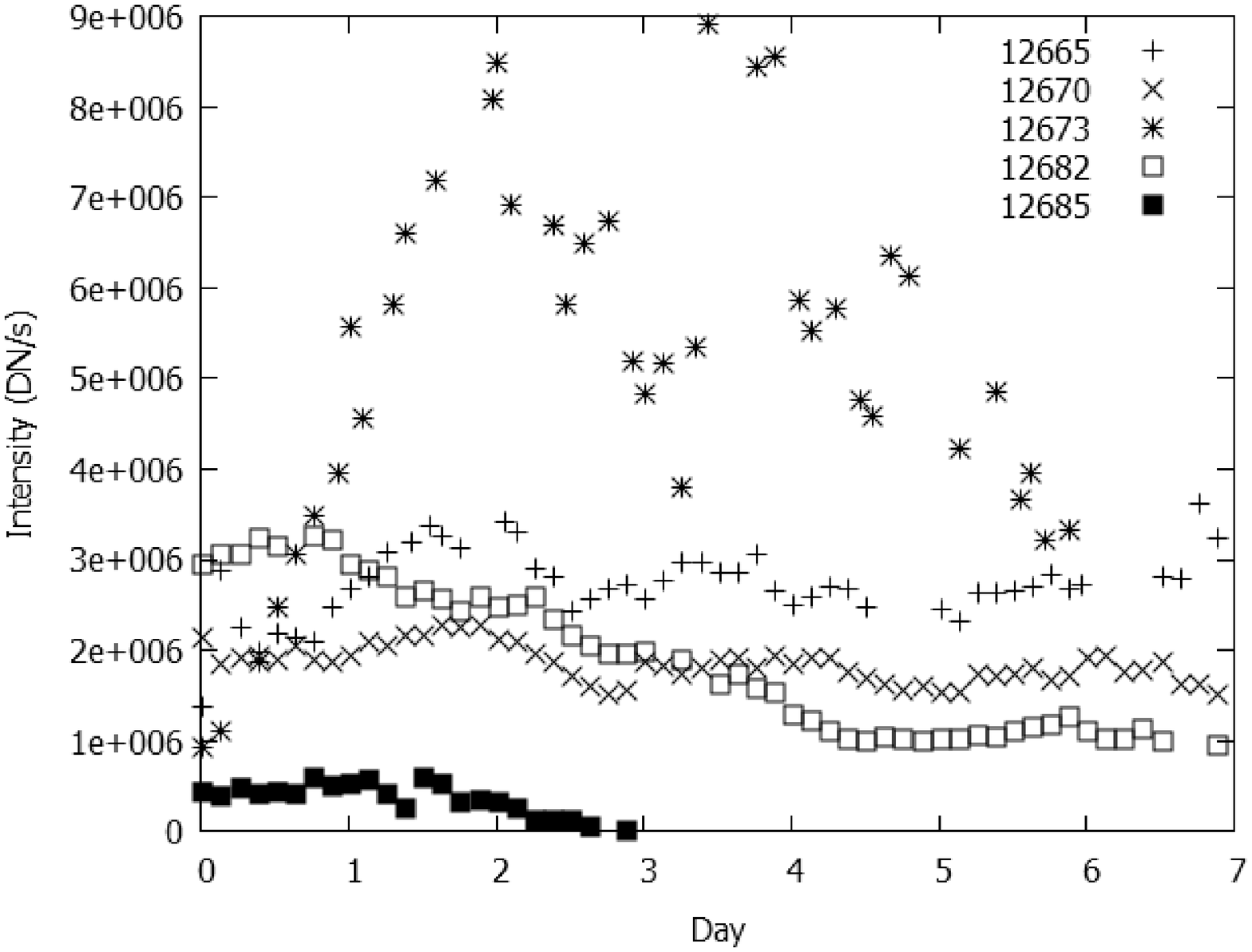}
\\
Fig.\ 13 AIA 304 $\mathrm{\AA}$ total intensities of AR 12665 starting from July 8, AR 12670 starting from August 4, AR 12673 starting from September 3, AR 12682 starting from September 26, and AR 12685 starting from October 23, 2017.
\\

\end{document}